\documentclass[letterpaper]{achemso}

\usepackage[T1]{fontenc}

\usepackage{geometry}
\usepackage{setspace}

\usepackage{graphicx}
\usepackage{float}
\newfloat{scheme}{htbp}{los}
\floatname{scheme}{Scheme}
\floatname{chart}{Chart}
\newfloat{graph}{htbp}{loh}

\usepackage{chemformula} 
\usepackage[version = 4]{mhchem} 
\usepackage{braket}
\usepackage{graphicx}
\usepackage{amsmath}
\usepackage{booktabs}
\usepackage{amsmath}
\usepackage{adjustbox}
\usepackage{multirow}
\usepackage{float}
\usepackage{array}
\usepackage{arydshln}
\usepackage{tabularx} 
\usepackage{multirow}
\usepackage{arydshln}
\newcommand{\abinitio}{{\textit{ab initio }}}
\newcommand{\Be}{B_{\rm e}}

\newcommand{\dprime}{\prime \prime}

\newcommand*{\Nocc}{N_{\rm occ}}

\newcommand*{\low}{{\textrm{L\"owdin}}}
\newcommand*{\LOlow}{LO(L\"owdin)}
\newcommand*{\NEO}{$N_{\rm EO}$}
\newcommand{\SI}[1]{{\color{blue} #1}}

\newcommand*{\orth}[1]{ {\rm orth}\left(#1 \right)}

\title{Impurity-Preserved Density Matrix Embedding Theory for Local Electronic Excitations}
\author{Teng Zhang}
\author{Ze-Wei Li}
\author{Zhe-Bin Guan*}
\author{Hong Jiang}
\date{\today}
\email{jianghchem@pku.edu.cn}
\affiliation[Peking University]
{Beijing National Laboratory for Molecular Sciences, Institute of Theoretical and Computational Chemistry, College of Chemistry and Molecular Engineering, Peking University, Beijing 100871, China}

\begin{document}

\maketitle
\newpage

\begin{abstract}
Density matrix embedding theory (DMET), which is usually based on a Schmidt decomposition of Slater determinants by partitioning the full system into impurity and environment in terms of local orthogonal orbitals (LOs), has demonstrated considerable promise in electronic structure studies because it enables the extraction of local properties using a high-level solver within an embedded impurity subsystem with greatly reduced degrees of freedom, thereby achieving a balance between accuracy and computational cost. However, its application to excited states of strongly correlated systems, such as lanthanide complexes, remains challenging because the errors relative to all-electron results can still be significant. Motivated by the success of the previously developed atomic orbitals (AOs) based DMET framework (Ai, Li, and Jiang, Phys. Rev. Lett. 2025, 135, 026502.), termed AO-DMET, which attains improved accuracy by constructing the embedded subspace based on a non-orthogonal decomposition of the Slater determinant in terms of AOs, we propose a new LO-based partitioning scheme that fully preserves the impurity space spanned by corresponding AOs and can achieve accuracy closely matching that of AO-DMET while retaining the orthogonal partition and its associated computational efficiency. The performance of the proposed method is demonstrated through excitation energy calculations for several representative lanthanide complexes. These results establish an efficient and accurate partitioning scheme for describing excited states in strongly correlated systems within the DMET framework.
\end{abstract}
\clearpage

\section{INTRODUCTION}

Over the years, quantum embedding methods have become powerful tools for reducing the computational cost of high-level quantum chemistry methods for large strongly correlated systems \cite{Sun2016, Jones2020, Vorwerk2022, Verma2026}. By assuming that the target properties of the full system are predominated by strong electronic correlations within a specific region, termed ``impurity'' and denoted as $A$ henceforth, quantum embedding methods convert the high-level treatment of the full Hamiltonian to that of an embedded impurity Hamiltonian with greatly reduced degrees of freedom. The latter should approximately account for quantum many-body interactions, also termed entanglement \cite{Peschel2012}, between $A$ and its environment $B$ that are extracted from a low-level, e.g. Hartree-Fock (HF) or density-functional theory (DFT), treatment of the whole system. Among various quantum embedding approaches that have been developed in recent decades, including those based on electron density \cite{Libisch2014, Wesolowski2015, Jacob2024}, density matrix \cite{Knizia2012, Welborn2016, Fornace2015, Yu2017, Lau2021, Verma2026} and Green's function \cite{Zgid2011, Lan2015, Ma2021, Zhu2021}, density-matrix embedding theory (DMET) developed by Chan and coworkers \cite{Knizia2012,Knizia2013,Wouters2016} has recently demonstrated strong potential in applications for systems characterized by local excitations and/or correlations such as point defects in solids \cite{Haldar2023, Mitra2021, Mitra2023, Verma2023, Lau2024, Otis2025} and single-ion magnets (SIMs) \cite{Ai2022, Ai2025a,GuanZB2025, Huang2025}, owing to its relatively straightforward and mathematically rigorous framework, favorable computational scaling, and systematically improvable accuracy. DMET can also be formulated as a local correlation strategy \cite{Saebo1993} that allows efficient treatment of large molecular systems or periodic solids by high-level quantum chemistry solvers \cite{Cui2020b, LiC2023, Pham2020, Cui2022, Nusspickel2022, SunY2024, SunY2025}, and it has also inspired the development of other highly promising embedding approaches \cite{Welborn2016, Ye2018, Fertitta2019, Scott2024, Cernatic2024, LiJ2024, Ai2025b, Makhlouf2025}.

The standard DMET formalism requires partitioning the full system into the impurity $A$ and environment $B$ in terms of a set of localized orthogonal orbitals (LOs) that can be uniquely assigned to either $A$ or $B$. There are various schemes for LO construction, roughly falling into two categories named as \textit{top-down} and \textit{bottom-up}, respectively \cite{Cui2020b}. LOs in the \textit{top-down} category are constructed by using some localization transformation of mean-field canonical molecular orbitals (CMOs), such as intrinsic and projected atomic orbitals (IAO+PAO) \cite{Knizia2013IAO, Cui2020b}, Boys \cite{Foster1960}, Pipek-Mezey (PM) \cite{Pipek1989} and Edmiston-Ruedenberg \cite{Edmiston1963} localized MOs (LMOs), and maximally localized Wannier functions (MLWF) \cite{Marzari2012} for periodic systems. LOs in the \textit{bottom-up} category are constructed by performing some orthogonalization transformation of atomic orbital (AO) basis functions without using any information of CMOs, such as L\"owdin \cite{Loewdin1950} and meta-L\"owdin orthogonalization \cite{Sun2014}. LOs from both categories have been used in previous DMET studies. For example, IAO+PAO were used by Chan and coworkers in $\abinitio$ DMET studies of molecular and periodic systems \cite{Wouters2016, Cui2020b}, and meta-$\low$ LOs were used by Gagliardi and co-workers in DMET-based complete active space self-consistent field (CASSCF) studies of transition-metal complexes \cite{Verma2025}, and $\low$ LOs were used in our previous DMET-based CASSCF studies of 3d and 4f single-ion magnets \cite{Ai2022, Ai2025a, GuanZB2025}. The effects of the choice of LOs on the performances of DMET have been rarely addressed to the best of our knowledge. A recent DMET-based CCSD(T) study of water clusters by Sun \cite{SunY2025} found that using $\low$ LOs in DMET leads to significantly better agreement with all-electron results compared to IAO+PAO based DMET.

Recently, we developed a novel DMET formulation based on a non-orthogonal decomposition of Slater determinants in which the full system is partitioned into the impurity ($A$) and environment ($B$) in terms of  non-orthogonal atomic orbitals (AOs), hence termed AO-DMET \cite{Ai2025a}. Besides eliminating the arbitrariness in the construction of LOs, AO-DMET exhibits significantly improved accuracy for electronic excitation properties of transition metal and lanthanide complexes compared to the original LO-DMET with $\low$ LOs. In this work, we propose a new LO construction scheme inspired by the formalism of AO-DMET that can essentially reproduce the accuracy of AO-DMET for the description of local excitations on the one hand, and maintain formal and practical advantages of the original LO-DMET on the other hand.

The paper is organized as follows. First, the basic formalisms of both LO- and AO-DMET are reviewed as necessary backgrounds. We then present the new LO construction scheme and prove in a mathematically rigorous manner that the embedded impurity space of LO-DMET with the new LO construction is exactly a subspace of that of AO-DMET. To demonstrate the performance of the proposed method, excitation energies of four lanthanide complexes, taken as representative strongly correlated systems, are calculated using AO-DMET and LO-DMET with different LO constructions. Finally, we summarize the main conclusions and discuss possible directions for further development.

\section{METHODS AND COMPUTATIONAL DETAILS}

\subsection{Localized Orthogonal Orbitals Based Density Matrix Embedding Theory}

The basic formalism of density matrix embedding theory based on localized orthogonal orbitals (LOs) is briefly reviewed here following ref. \citenum{Wouters2016}. We begin with a set of atom-centered LOs $\{\phi_{\mu}\}$ for the total system, which are partitioned into $N_A$ impurity LOs $\{\phi^A_\alpha\}$ and $N_B$ environment LOs $\{\phi^B_\beta\}$. Occupied canonical molecular orbitals (MOs) of the ground state Slater determinant $\Phi_0$ can then be expanded as
\begin{equation} \label{eq:chi}
	|\chi_i\rangle=\sum_{\alpha \in A}C_{\alpha i}^A|\phi_{\alpha}^A\rangle+\sum_{\beta \in B}C_{\beta i}^B|\phi_{\beta}^B\rangle.
\end{equation}
Without loss of generality, we assume $N_A < \Nocc < N_B$, with $\Nocc$ being the number of occupied MOs, which usually holds in most cases, and the requirement can be relieved in practice. Applying the singular value decomposition (SVD) to the impurity block of the expansion coefficient matrix (i.e. $C_{\alpha i}^A$) in eq. \eqref{eq:chi} leads to
\begin{equation}
  \mathbf{C}^{A} = \mathbf{U \Sigma V^\dagger},
\end{equation}
where $\mathbf{U}$ and $\mathbf{V}$ are $N_A \times N_A$ and $\Nocc \times \Nocc$ unitary matrices, respectively, and $\mathbf{\Sigma}$ is the diagonal matrix of $N_A \times \Nocc$ with $\Sigma_{\alpha i}=\sigma_i \delta_{\alpha, i}$. Applying a unitary transform of occupied MOs with $\mathbf{V}$, one can obtain \cite{Wouters2016}
\begin{equation} \label{eq:chi-rotated}
|\tilde{\chi}_i\rangle = \sum_jV_{ji}|\chi_j\rangle =
\left\{
   \begin{matrix}
		 \sigma_i |A_i \rangle+\sqrt{1-\sigma_i^2}|B_i\rangle &  \quad  (1 \le i \le N_A) \\
		  \sum_{\beta, j}  C_{\beta j}^B V_{ji} \ket{\phi_\beta^B}     &  \quad  (i > N_A)
	\end{matrix}
\right.,
\end{equation}
where
\begin{eqnarray}
  |A_i\rangle &=& \sum_{\alpha}U_{\alpha i}|\phi_{\alpha}^{A}\rangle \\
  |B_i\rangle &=& \sum_{\beta}Y_{\beta i}|\phi^B_\beta\rangle,
\end{eqnarray}
with
\begin{equation} \label{eq:Y}
   Y_{\beta i}=\sum_{j}V_{ji} C_{\beta j}^B/\sqrt{1-\sigma_i^2}.
\end{equation}
Eq. \eqref{eq:chi-rotated} indicates that among $\Nocc$ rotated occupied orbitals, at most $N_A$ of them have contributions in both $A$ and $B$, and the associated $\{\ket{B_i}\}$, denoted as $\Be$, are defined as bath orbitals, which describe the entanglement between $A$ and $B$ as embodied in $\ket{\Phi_0}$ \cite{Knizia2012, Wouters2016}. The remaining $\ket{\tilde{\chi}_i}$ reside purely in $B$, and form unentangled occupied orbitals, usually termed core orbitals in the DMET literature \cite{Wouters2016}. An equivalent but more straightforward procedure \cite{Wouters2016} to construct bath and core orbitals is to diagonalize the environment block of the first-order reduced density matrix (1-RDM) of $\Phi_0$ in the LO basis. Among the resulting eigenvectors, those with eigenvalues $\lambda = 1$ correspond to core orbitals and those with eigenvalues $\lambda \in (0,1)$ correspond to the bath orbitals. Alternatively, one can also construct bath and core orbitals by SVD of the off-diagonal block of 1-RDM \cite{Cui2020b} or Householder transformation \cite{Sekaran2021}.

Using eq. \eqref{eq:chi-rotated}, one can obtain the Schmidt decomposition of $\ket{\Phi_0}$
\begin{equation} \label{Phi0-decomp-LO}
	\ket{\Phi_0} = \sum_{I}^{2^{N_A}} \Lambda_{I} \ket{\Phi_{I}^{A}}\otimes\ket{\Phi_{I}^{\Be}}
	\otimes \ket{\Phi^{\rm core}}= \ket{\Psi^{A+\Be}} \otimes \ket{\Phi^{\rm core}},
\end{equation}
where $\Phi_I^{A}$ and $\Phi_I^{\Be}$ denote the many-body basis functions in the Fock spaces spanned by impurity ($A$) and bath ($\Be$) orbitals, respectively, and $\Phi^{\rm core}$ denotes the Slater determinant formed by core orbitals. The combination of impurity and bath orbitals defines the embedded impurity orbitals (EO). By projecting the full Hamiltonian to the many-body space spanned by $\set{\ket{\Phi_I^{A}}\otimes\ket{\Phi_J^{\Be}}\otimes\ket{\Phi^{\rm core}}}$, we obtain the following embedded impurity Hamiltonian \cite{Wouters2016}
\begin{equation}
	\hat{H}_{\rm emb} = \sum_{p,q\in {\rm EO}}\tilde{h}_{pq}\hat{a}_{p}^{\dagger}\hat{a}_{q} +
	\frac{1}{2}\sum_{p,q,r,s\in {\rm EO}} \langle pq|rs\rangle \hat{a}_{p}^{\dagger} \hat{a}_{q}^{\dagger}\hat{a}_{s}\hat{a}_{r}+ E_{\rm core}.
\end{equation}
Here $\tilde{h}_{pq}$ are matrix elements of the following single-particle operator
\begin{equation} \label{eq:h}
\hat{\tilde{h}}=-\frac{1}{2}\nabla^2-\sum_I\frac{Z_I}{|\mathbf{r}-\mathbf{R}_I|}+\sum_{ a \in \mathrm{core}}(\hat{J}_a-\hat{K}_a),
\end{equation}
where $I$ runs over all nuclei in the full system $A+B$, with nuclear charge $Z_I$ and coordinate $\mathbf{R}_I$, and $\hat{J}_a$ and $\hat{K}_a$ denote the Coulomb and exchange operators generated by core orbital $a$. $E_{\rm core}$ denotes the HF energy corresponding to $\Phi^{\rm core}$.

\subsection{Non-orthogonal atomic orbitals based DMET}
As a generalization of the standard DMET formalism presented above, AO-DMET is formulated based on partitioning the full system into the impurity $A$ and environment $B$ in terms of non-orthogonal atomic orbitals (AOs), e.g. Gaussian basis functions widely used in most quantum chemistry calculations, denoted as $\{\ket{\phi^\prime_\mu}\}$. In this formulation, bath orbitals, accounting for both the entanglement and overlapping between $A$ and $B$, can be obtained by the following procedure:
\begin{enumerate}
 \item We first orthogonalize AOs centered within $B$ locally, using, e.g. $\low$ orthogonalization
\begin{equation} \label{eq:B-local-ortho}
   |\phi^{\prime \prime B }_\beta \rangle  = \sum_{\beta'} X_{\beta' \beta}^{B} |\phi^{\prime B}_{\beta'} \rangle,
\end{equation}
where $\mathbf{X}^B=\mathbf{S}_B^{-1/2}$ with $\mathbf{S}_B$ being the overlap matrix corresponding to AOs within $B$.
 \item We then expand occupied MOs in $|\Phi_0\rangle$ by using original AOs within $A$ and locally orthogonalized AOs within $B$,
\begin{equation}
	|\chi_i\rangle=\sum_\alpha C^{\prime A}_{\alpha i}|\phi_\alpha^{\prime A}\rangle+\sum_\beta C^{\dprime B}_{\beta i}|\phi_{\beta}^{\dprime B}\rangle,
	\label{AOMO}
\end{equation}
and perform the SVD of the environment block of the expansion matrix $\mathbf{C}^{\dprime B}$ as
\begin{equation} \label{eq:CB-SVD}
  C^{\dprime B}_{\beta i}=\sum_j U_{\beta j}^{\dprime}\sigma^{\dprime}_j[V^{\dprime}]_{ji}^\dagger.
\end{equation}
Under the assumption $ 2 N_A < \Nocc < N_B$, we can prove\cite{Ai2025b} that among $\Nocc$ singular values $\sigma^{\dprime}_j$, at most $2 N_A$ of them will have values either $0 < \sigma^{\dprime}_j < 1$ or  $\sigma_j^{\dprime}>1$, and the remaining ones will be exactly equal to 1.
\item We then rotate the occupied MOs with $\mathbf{V}^{\dprime}$
\begin{equation}
	\ket{\tilde{\chi}_i}
	=\sum_jV_{ji}^{\dprime} \ket{\chi_j}
	=\sum_{\alpha j} C^{\prime A}_{\alpha j}V_{ji}^{\dprime}\ket{\phi^{\prime A}_\alpha}
	+\sigma_i^{\dprime} \ket{B_i^{\dprime}},
\end{equation}
with
\begin{equation}
	\ket{B_i^{\dprime}}=\sum_\beta U_{\beta i}^{\dprime} \ket{\phi_\beta^{\dprime B}}.
\end{equation}
It can be proved \cite{Ai2025b} that $\ket{\tilde{\chi}_i}$ with $\sigma_i^{\dprime}=1$ contain contributions only from the environment, i.e. $\ket{\tilde{\chi}_i} = \ket{B_i^{\dprime}} $, and therefore these are occupied orbitals purely residing in the environmental region, identified as ``core orbitals'', similar to those defined in LO-DMET. The remaining $\ket{\tilde{\chi}_i}$ with either $0 < \sigma^{\dprime}_i < 1$ or  $\sigma^{\dprime}_i>1$ have contributions from both the impurity and environment, and the corresponding $\ket{B_i^{\dprime}}$ are identified as ``bath orbitals'', whose number is at most $2 N_A$, and they account for both the entanglement (in the same sense as that in LO-DMET) and overlapping between $A$ and $B$ as embodied in $\ket{\Phi_0}$, therefore denoted as $B_{\rm eo}$ henceforth.
\end{enumerate}
One can further prove that the core orbitals are exactly orthogonal to both bath orbitals and AOs within $A$ \cite{Ai2025b}, and therefore the Slater determinant $\ket{\Phi_0}$ can be decomposed as
\begin{equation} \label{eq:Phi0-decomp-AO}
 \ket{\Phi_0} = \ket{\Psi^{A + B_{\rm eo}}} \otimes\ket{{\bar{\Phi}}^{\mathrm{core}}},
\end{equation}
where $\ket{\Psi^{A + B_{\rm eo}}}$ is a many-body wavefunction defined in the space that is spanned by the union of AOs within $A$ and bath orbitals in $B_{\rm eo}$. The corresponding embedded impurity Hamiltonian reads
\begin{equation}
	\hat{H}_{\rm emb}^{\rm AO-DMET} = \sum_{p,q\in A+B_{\rm eo}}\bar{h}_{pq}\hat{a}_{p}^{\dagger}\hat{a}_{q}+\frac{1}{2}\sum_{pqrs \in A+B_{\rm eo}}\bar{V}_{pqrs}\hat{a}_{p}^{\dagger}\hat{a}_{q}^{\dagger}\hat{a}_{s}\hat{a}_{r} + {\bar{E}}_{\rm core}
	\label{eq:Hemb-AO}
\end{equation}
with
\begin{eqnarray}
  \bar{h}_{pq}   &\equiv& [\mathbf{S}^{-1}\mathbf{\tilde{h}}\mathbf{S}^{-1}]_{pq} \\
  \bar{V}_{pqrs} &\equiv& \sum_{p'q'r's'\in A+B_{\rm eo} }S_{pp^{\prime}}^{-1}S_{qq^{\prime}}^{-1}S_{rr^{\prime}}^{-1}S_{ss^{\prime}}^{-1}\langle p^{\prime}q^{\prime}|r^{\prime}s^{\prime}\rangle,
\end{eqnarray}
where $\mathbf{S}$ is the overlap matrix and $\mathbf{\tilde{h}}$ is the matrix of the single-particle operator given in Eq. \eqref{eq:h} between orbitals in $A+B_{\rm eo}$ (i.e. EOs in AO-DMET). Similar to that in LO-DMET, bath and core orbitals in AO-DMET can also be obtained by diagonalizing the environment block of 1-RDM expressed in the basis $\{{\phi}_{\beta}^{\dprime B}\}$, $[\mathbf{D}^{\dprime}]^{BB}$, but with a slightly different classification criterion: eigenvectors with $\lambda = 1$ are still identified as core orbitals, whereas those with $\lambda \in (0,1)\cup(1,\infty)$ are identified as bath orbitals. Further details can be found in ref.~\citenum{Ai2025b}.

Compared to LO-DMET, AO-DMET exhibits several distinct advantages \cite{Ai2025b}. For example, it avoids the extended tails of LOs, and yields bath orbitals with stronger locality. The latter allows the possibility of introducing a small auxiliary basis to expand bath orbitals that can reduce the computational cost related to the calculation of two-electron integrals when considering spin-orbit coupling. It can be more efficient in certain cases, such as when a fragmentation approximation is applicable. In particular, excitation energy calculations for strongly correlated systems, including transition-metal SIMs and lanthanide luminescent complexes, indicate that AO-DMET yields results in significantly better agreement with the corresponding all-electron reference values than LO(L\"owdin)-DMET, suggesting its improved ability to capture local correlation effects. A minor drawback of AO-DMET is that, for the same choice of $A$, it typically yields more bath orbitals than LO-DMET, leading to a larger EO space and consequently higher computational cost. Specifically, one can prove that under the condition of $N_{\rm occ} \geq 2N_A$, the number of bath orbitals in AO-DMET is at most $2 N_A$\cite{Ai2025b}, while that in LO-DMET is at most $N_A$, as mentioned previously.

\subsection{DMET based on impurity preserved orthogonal local orbitals}

In spite of the advantages of AO-DMET discussed above, the orthogonal partitioning underlying LO-DMET is still conceptually appealing. Considering the non-uniqueness of obtaining LOs, it is of great interest to explore the possibility of constructing LOs in such a way that the subsequent LO-DMET would have comparable performances of AO-DMET. We found that by using LOs constructed in a manner inspired by the formalism of AO-DMET, we can obtain a new implementation of LO-DMET that can achieve essentially the same accuracy of AO-DMET. The new scheme takes the following procedure. One conducts an orthogonalization of AOs within $A$ first by, e.g., $\low$ orthogonalization, yielding $\{\phi^{A}_{\alpha}\}=\orth{\{\phi^{\prime A}_{\alpha^{\prime}}\}}$. Defining the projector
\begin{equation}
	\hat{Q}_A=\hat{I}- \sum_\alpha \ket{\phi_{\alpha}^{A}} \bra{\phi_{\alpha}^{A}},
\end{equation}
one obtains LOs corresponding to $B$, denoted as $\ket{\phi_\beta^{B}}$, by applying $\hat{Q}_A$ to all AOs within $B$ and subsequently performing an orthogonalization on them,
\begin{equation}
	\{\ket{\phi_\beta^{B}}\} =\orth{\{ \hat{Q}_A\ket{\phi_{\beta^{\prime}}^{\prime B}}\}}.
\end{equation}
 
Obviously, LOs of $A$ span the same space as AOs of $A$, and this construction scheme is therefore referred to as ``impurity-preserved'' (IP). In addition, an important property of the IP construction can be established: the EO space of LO(IP)-DMET forms a subspace of the EO space in the AO-DMET when using the same $A$-$B$ division. \SI{A detailed proof is provided in sec.S1 in the Supporting Information}.

To showcase the differences between LO-DMET based on different LOs, we will compare the results from LO(IP)-DMET to those obtained from DMET based on L\"owdin LOs and IAO+PAO \cite{Knizia2013IAO, Cui2020b}, both of which have been widely used in previous DMET studies as mentioned in the Introduction section. \SI{Some details on the construction of IAO+PAO used in this work are provided in sec. S2 in the Supporting Information}.

\subsection{Computational Details}

\begin{figure}[H]
	\centering
	\includegraphics[width=\linewidth]{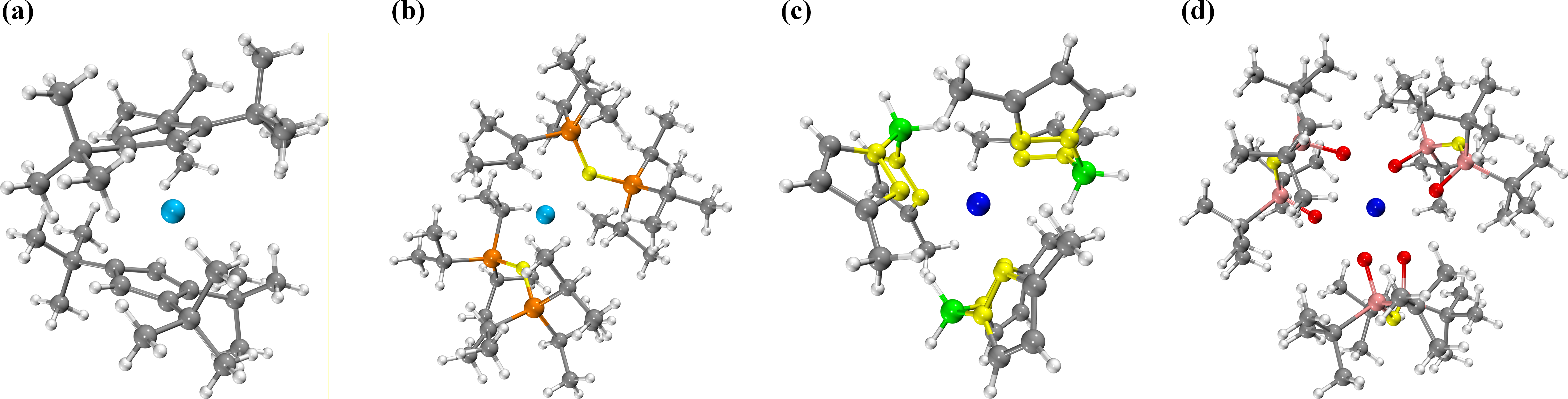}
	\caption{Structures of four lanthanide molecules used for tests in this work,
		(a) [Dy(Cp$^{\mathrm{ttt}}$)$_2$]$^+$ ($\textbf{1Dy}$ for short), 
		(b) [$\mathrm{Dy\{N[Si({}^{i}Pr)_3][Si({}^{i}Pr)_2C(CH_3)=CHCH_3]\}}$]$^+$ ($\textbf{2Dy}$ for short), 
		(c) Ce-Bp$^{\mathrm{Me}}$ ($\textbf{3Ce}$ for short), and
		(d) Ce-O$_2$ip$^{\mathrm{tBu}}$ ($\textbf{4Ce}$ for short),
		obtained from Refs.~\citenum{Goodwin2017},\citenum{Emerson2025}, \citenum{GuoR2022} and \citenum{ZhengJ2025}, respectively. Silver/orange/white/green/yellow/blue/cyan/pink/red represent C/Si/H/B/N/Ce/Dy/P/O atoms, respectively.}
	\label{structures}
\end{figure}

To comprehensively test the performances of the LO(IP)-DMET approach for the description of local electronic excitations, we consider four lanthanide complexes, two Dy(III) complexes, denoted as \textbf{1Dy} and \textbf{2Dy}, that exhibit intriguing single-ion magnet (SIM) features \cite{Goodwin2017, Emerson2025}, and two Ce(III) complexes, denoted as \textbf{3Ce} and \textbf{4Ce}, that exhibit interesting luminescent properties featuring 4f-5d transitions, and their structures are illustrated in Fig.~\ref{structures}.

The basis sets employed are as follows. For \textbf{1Dy} and \textbf{2Dy}, ANO-RCC-VTZP \cite{Widmark1990,Roos2004,Roos2008} is used for Dy and its nearest coordinating atoms (ten C atoms for \textbf{1Dy}, and N together with two C atoms of the pendant alkene for \textbf{2Dy}); all remaining atoms in \textbf{1Dy} are described using the ANO-RCC-VDZP \cite{Widmark1990, Roos2004, Roos2008}, while in \textbf{2Dy}, ANO-RCC-VDZP is used for Si and the remaining C atoms, and ANO-R0 \cite{Widmark1990,Zobel2019,Zobel2021} is used for H. For \textbf{3Ce} and \textbf{4Ce}, ANO-RCC-VTZP is used for Ce, O and N in \textbf{3Ce}; ANO-RCC-VDZP for B, P and N in \textbf{4Ce}, and ANO-R0 for C and H.

In LO/AO-DMET, bath and core orbitals are obtained by diagonalizing the environment block of the 1-RDM in the LOs/AOs representation \cite{Wouters2016, Ai2025b} obtained from the restricted open-shell HF (ROHF) calculation of the full system. A threshold parameter $\varepsilon_{\rm bath}$ for distinguishing eigenvalues in different ranges is uniformly set to $10^{-12}$\cite{Ai2022,Ai2025b}.

For the high-level solver, we use the state-averaged (SA) CASSCF to treat static correlation and consider dynamical correlation correction by the strongly contracted second-order $n$-electron valence state perturbation theory (SC-NEVPT2) \cite{Angeli2001, Angeli2001CPL}. Scalar relativistic effects are treated using the one-electron variant of the spin-free exact two-component theory (SFX2C-1e)\cite{Kutzelnigg2005, Liu2009, Dyall2001}. Spin--orbit coupling (SOC) effects are incorporated through the state-interaction spin--orbit (SISO) method \cite{Atanasov2015, Malmqvist2002, Sayfutyarova2016}, in which the spin-dependent SOC term is approximated by the spin--orbit mean-field (SOMF) approximation to the Breit--Pauli Hamiltonian \cite{Neese2005,Hess1996}. For SA-CASSCF calculations, all 21 sextet states in the CAS(9e,7o) comprising Dy 4f orbitals are used for \textbf{1Dy} and \textbf{2Dy}, and 12 doublets in the CAS(1e,12o) comprising Ce 4f and 5d orbitals are considered for \textbf{3Ce} and \textbf{4Ce}. All calculations were carried out using our local extension of the PySCF package \cite{Sun2020}.

\section{RESULTS AND DISCUSSION}

In this section, we compute the excitation energies of four test molecules within the DMET framework using different partitioning schemes, including AO-DMET, LO(L\"owdin)-DMET, LO(IAO+PAO)-DMET, and the newly proposed LO(IP)-DMET. Since one of the key challenges in DMET treatments of lanthanide complexes is capturing strong correlation effects with a limited EO space, a comprehensive assessment of the methods above requires analyzing their ability to describe correlation at different levels.

To examine the treatment of static correlation, we present excitation energies obtained from CASSCF calculations. To further assess the ability to capture dynamical correlation, excitation energies corrected by subsequent NEVPT2 calculations are also reported. For both methods, SOC effects are taken into account in the commonly used two-step approach. In all calculations, the impurity $A$ is chosen to be either the central lanthanide ion alone or the lanthanide ion together with its nearest coordinating atoms, as these choices are consistent with chemical intuition and are commonly adopted in DMET calculations \cite{Ai2022,Ai2025a,Ai2025b,GuanZB2025}.

\begin{table}[H]
\centering
\caption{Excitation energies (in cm$^{-1}$) of seven lowest excited-state Kramers doublets (KDs) with respect to the ground-state KD in \textbf{1Dy} obtained from all-electron (AE) and different DMET variants based CASSCF and NEVPT2. The choices of impurity ($A$) are indicated on the second row, with the corresponding numbers of impurity orbitals given in the parentheses. The third row shows the numbers of embedded impurity orbitals ($N_{\mathrm{EO}}$) in different treatments. Mean absolute error (MAE) and mean absolute relative error (MARE) are evaluated with respect to the corresponding AE results shown in the last column.} \label{tab:1Dy}
\resizebox{\textwidth}{!}{%
\begin{tabular}{lccccccccc}
\toprule
Method& AO   & LO(IP) & LO(L\"owdin) & LO(IAO+PAO) & AO  & LO(IP) & LO(L\"owdin) & LO(IAO+PAO) & AE \\
	\midrule
$A$    & \multicolumn{4}{c}{Dy(104)}               & \multicolumn{4}{c}{DyC$_{10}$(404)}       & \\
	\cmidrule(lr){2-5} \cmidrule(lr){6-9}
	$N_{\mathrm{EO}}$ & 270 & 213 & 213 & 140 	& 570 & 570 & 570 & 483 & 1030 \\
			\midrule
			\multicolumn{10}{l}{\textbf{CASSCF}} \\
			\midrule
			1 & 487.9  & 488.0  & 490.0  & 541.7  & 486.4  & 486.4  & 486.6  & 491.4  & 486.2 \\
			2 & 772.5  & 772.6  & 776.4  & 874.7  & 769.2  & 769.2  & 769.6  & 778.9  & 768.8 \\
			3 & 954.4  & 954.5  & 959.4  & 1084.5 & 949.8  & 949.8  & 950.5  & 962.7  & 949.4 \\
			4 & 1117.6 & 1117.7 & 1123.1 & 1260.3 & 1112.5 & 1112.5 & 1113.2 & 1126.9 & 1111.9 \\
			5 & 1274.2 & 1274.3 & 1280.1 & 1422.6 & 1268.9 & 1268.9 & 1269.6 & 1284.3 & 1268.3 \\
			6 & 1398.4 & 1398.5 & 1404.2 & 1542.1 & 1393.2 & 1393.2 & 1394.0 & 1408.4 & 1392.6 \\
			7 & 1482.1 & 1482.3 & 1489.0 & 1663.0 & 1475.5 & 1475.5 & 1476.5 & 1495.0 & 1474.7 \\
			\hline
			MAE & 5.0 & 5.1 & 10.1 & 133.9 & 0.5 & 0.5 & 1.2 & 13.7 & \\
			MARE (\%) & 0.5 & 0.5 & 0.9 & 12.6 & 0.05 & 0.05 & 0.1 & 1.3 & \\
			\midrule
			\multicolumn{10}{l}{\textbf{NEVPT2}} \\
			\midrule
			1 & 491.9  & 492.0  & 533.4  & 600.3  & 509.3  & 509.3  & 512.7  & 525.2  & 507.5 \\
			2 & 741.0  & 741.1  & 808.7  & 952.0  & 765.6  & 765.6  & 772.4  & 790.3  & 762.3 \\
			3 & 904.2  & 904.4  & 980.8  & 1178.2 & 925.5  & 925.5  & 934.6  & 951.4  & 921.4 \\
			4 & 1084.3 & 1084.5 & 1162.8 & 1390.5 & 1097.7 & 1097.7 & 1108.4 & 1121.7 & 1093.2 \\
			5 & 1277.8 & 1277.8 & 1356.1 & 1600.5 & 1282.0 & 1282.0 & 1293.7 & 1303.8 & 1277.2 \\
			6 & 1436.8 & 1436.8 & 1519.1 & 1762.8 & 1439.1 & 1439.1 & 1451.6 & 1459.5 & 1434.0 \\
			7 & 1561.5 & 1561.6 & 1628.8 & 1912.5 & 1541.5 & 1541.6 & 1553.1 & 1560.6 & 1536.0 \\
			\hline
			MAE & 13.1 & 13.1 & 65.5 & 266.5 & 4.2 & 4.2 & 13.6 & 25.8 & \\
			MARE (\%) & 1.5 & 1.5 & 6.0 & 24.4 & 0.4 & 0.4 & 1.3 & 2.6 & \\
			\bottomrule
		\end{tabular}%
	}
\end{table}

\begin{table}[H]
	\centering
	\caption{Excitation energies (in cm$^{-1}$) of seven lowest excited-state Kramers doublets (KDs) with respect to the ground-state KD in \textbf{2Dy} obtained from all-electron (AE) and different DMET variants based CASSCF and NEVPT2. The choices of impurity ($A$) are indicated on the second row, with the corresponding numbers of impurity orbitals given in the parentheses. The third row shows the numbers of embedded impurity orbitals ($N_{\mathrm{EO}}$) in different treatments. Mean absolute error (MAE) and mean absolute relative error (MARE) are evaluated with respect to the corresponding AE results shown in the last column. }
	\label{tab:2Dy}
	\resizebox{\textwidth}{!}{%
		\begin{tabular}{lccccccccc}
			\toprule
			Method& AO   & LO(IP) & LO(L\"owdin) & LO(IAO+PAO) & AO  & LO(IP) & LO(L\"owdin) & LO(IAO+PAO) & AE \\
			\midrule
			$A$ 
			& \multicolumn{4}{c}{Dy(104)}
			& \multicolumn{4}{c}{DyN$_2$C$_2$(224)}
			& \\
			\cmidrule(lr){2-5} \cmidrule(lr){6-9}
			$N_{\mathrm{EO}}$ 
			& 301 & 213 & 213 & 141
			& 443 & 443 & 443 & 281
			& 854 \\
			\midrule
			\multicolumn{10}{l}{\textbf{CASSCF}} \\
			\midrule
			1 & 560.9  & 561.0  & 563.0  & 633.4  & 559.4  & 559.4  & 559.7  & 578.7  & 558.9 \\
			2 & 1089.1 & 1089.2 & 1093.5 & 1216.3 & 1086.2 & 1086.2 & 1086.8 & 1118.4 & 1085.3 \\
			3 & 1508.4 & 1508.5 & 1514.9 & 1674.4 & 1504.4 & 1504.4 & 1505.1 & 1544.0 & 1503.1 \\
			4 & 1762.2 & 1762.3 & 1769.9 & 1942.6 & 1757.5 & 1757.5 & 1758.4 & 1798.4 & 1756.0 \\
			5 & 1902.1 & 1902.2 & 1910.1 & 2087.2 & 1897.2 & 1897.2 & 1898.2 & 1940.6 & 1895.6 \\
			6 & 2023.2 & 2023.3 & 2031.4 & 2219.4 & 2018.0 & 2018.0 & 2019.1 & 2064.1 & 2016.0 \\
			7 & 2081.2 & 2081.4 & 2089.3 & 2289.7 & 2075.7 & 2075.7 & 2076.9 & 2124.9 & 2073.4 \\
			\hline
			MAE & 5.5 & 5.6 & 12.0 & 167.8 & 1.4 & 1.4 & 2.3 & 40.1 & \\
			MARE (\%) & 0.35 & 0.36 & 0.77 & 11.15 & 0.09 & 0.09 & 0.14 & 2.71 & \\
			\midrule
			\multicolumn{10}{l}{\textbf{NEVPT2}} \\
			\midrule
			1 & 544.5  & 544.6  & 577.8  & 706.4  & 556.9  & 556.9  & 561.7  & 581.5  & 557.4 \\
			2 & 1092.0 & 1092.0 & 1150.3 & 1370.9 & 1109.3 & 1109.3 & 1117.7 & 1152.4 & 1109.0 \\
			3 & 1549.2 & 1549.2 & 1625.8 & 1909.8 & 1566.0 & 1566.0 & 1577.1 & 1621.6 & 1564.7 \\
			4 & 1835.3 & 1835.2 & 1921.0 & 2229.9 & 1848.7 & 1848.7 & 1861.6 & 1907.6 & 1846.6 \\
			5 & 1992.5 & 1992.4 & 2079.0 & 2394.8 & 2000.6 & 2000.6 & 2014.6 & 2062.5 & 1998.7 \\
			6 & 2130.3 & 2130.2 & 2215.6 & 2550.4 & 2131.9 & 2131.9 & 2145.9 & 2189.6 & 2129.6 \\
			7 & 2204.2 & 2204.1 & 2285.2 & 2636.2 & 2200.4 & 2200.4 & 2213.7 & 2247.2 & 2198.2 \\
			\hline
			MAE & 9.9 & 9.9 & 64.4 & 342.0 & 1.5 & 1.5 & 12.6 & 51.2 & \\
			MARE (\%) & 0.87 & 0.86 & 3.91 & 21.81 & 0.09 & 0.09 & 0.78 & 3.35 & \\
			\bottomrule
		\end{tabular}%
	}
\end{table}

\subsection{Lanthanide single-ion magnets}

We first consider two representative Dy(III) based lanthanide single-ion magnets (LnSIMs) that have attracted tremendous interest in the community of single-molecule magnetism \cite{Meng2016, Lunghi2022, Chilton2022, Wang2023, Vieru2024} because of their intriguing magnetic properties as a result of intricate interplay between strong electron correlation, spin-orbit coupling and crystal field splitting \cite{Rinehart2011}. The main quantity of interest for LnSIMs is the magnetic anisotropy characterized by some effective crystal field Hamiltonian with the corresponding parameters determined by the splitting of the ground state multiplet of lanthanide ions in different coordinating environments \cite{Rinehart2011, Chibotaru2023}. In this work, we will present the energies of seven lowest excited state Kramers doublets (KDs) with respect to the ground state KD obtained from the CASSCF or NEVPT2 based SISO treatment.

The CASSCF and NEVPT2 results for \textbf{1Dy} obtained from DMET with different partitioning schemes are presented in Table~\ref{tab:1Dy}, together with the results from all-electron treatment. We first compare the size of the embedded impurity space characterized by the number of EOs ($N_{\rm EO}$) as indicated in the third row. When Dy is chosen as the impurity, LO(IAO+PAO) yields the smallest number of EOs among the four partitioning schemes. This behavior can be attributed to the fact that occupied MOs can be exactly represented by IAOs, and therefore bath orbitals in LO(IAO+PAO)-DMET come purely from IAOs whose size is the same as the minimal atomic basis\cite{Knizia2013IAO}. EOs in the LO(IAO+PAO) scheme also include PAOs centered in the impurity region, and $N_{\rm EO}$ is at most the sum of the numbers of AOs and IAOs centered in the impurity region plus the number of open-shell MOs. It is also worth noting that LO(IP) yields the same number of EOs as LO(L\"owdin), whereas AO-DMET leads to significantly more EOs due to its non-orthogonal partition between $A$ and $B$. As the impurity $A$ is extended to \ce{DyC10}, the number of EOs in LO(IP), LO(L\"owdin) and AO-DMET becomes the same, and that in the LO(IAO+PAO) scheme is still significantly smaller. In this case, the computational disadvantage of AO-DMET relative to LO($\low$), arising from its non-orthogonal partitioning, is eliminated, and AO-DMET becomes equivalent to LO(IP)-DMET, following the previously proven relationship between their EO spaces.

In terms of accuracy, for the CASSCF calculations with Dy as the impurity, AO, LO(IP), and LO($\low$)-DMET produce results in good agreement with the AE values, with mean absolute errors (MAEs) around $10$ cm$^{-1}$ or less, which is an acceptable level of accuracy for the theoretical description of LnSIMs.\cite{Lunghi2022} In contrast, LO(IAO+PAO) shows significantly poorer accuracy, with the MAE of $133.9$ cm$^{-1}$. 
The accuracy can be significantly improved by including the nearest coordinating atoms in the impurity at the cost of increased computational expense, which is a common strategy in DMET \cite{Ai2022, Ai2025a, GuanZB2025, Mitra2021, Mitra2023, Haldar2023, Verma2023, Otis2025, Verma2025}. When treating \ce{DyC10} as the impurity, the errors of LO(IAO+PAO)-DMET can be dramatically reduced, with the MAE of 13.7 cm$^{-1}$, comparable to that of LO($\low$)-DMET treating Dy as the impurity. The former, however, has approximately twice as many EOs, which indicates that the less satisfactory performance of LO(IAO+PAO)-DMET is not merely due to the insufficient size of the EO space. Among all the partitioning schemes, LO(IP) exhibits the best overall performance, yielding the MAE of $5.1$ cm$^{-1}$ with the Dy-only impurity, nearly identical to that of AO-DMET but with 80\% as many EOs, and the MAE of $0.5$ cm$^{-1}$ with the \ce{DyC10} impurity. In both impurity choices, the errors of LO(IP)-DMET are approximately half of those of LO($\low$)-DMET.

For the NEVPT2 calculations with dynamical correlation taken into account, the errors of the excitation energies relative to the corresponding AE results increase for all impurity choices and partitioning schemes, indicating that dynamical correlation is more difficult to capture than static correlation within the DMET framework, which can be attributed to the stronger nonlocality of the former, as discussed in our previous study \cite{GuanZB2025}. Still, both LO(IP)- and AO-DMET show the MAE of $13.1$ cm$^{-1}$ when choosing Dy as the impurity, a superior performance compared to LO($\low$)- and LO(IAO+PAO)-DMET, which show MAEs of 65.5 and 266.5 cm$^{-1}$, respectively. When treating \ce{DyC10} as the impurity, the differences between LO(IP)/AO-DMET and all-electron results become marginal, with an MAE of only 4.2 cm$^{-1}$, and in this case LO($\low$)- and LO(IAO+PAO)-DMET also perform reasonably well.

To further assess the generality of the findings in \textbf{1Dy} presented above, we also examined \textbf{2Dy}, which features a coordination environment very different from that of \textbf{1Dy}. As shown in Table~\ref{tab:2Dy}, the overall performances of DMET based on different partitioning schemes are highly consistent with those observed in \textbf{1Dy}. These results suggest that the favorable performance of LO(IP)-DMET is robust for Dy SIMs with different coordination environments. Taken together, the results for \textbf{1Dy} and \textbf{2Dy} demonstrate that LO(IP)-DMET is a potentially efficient and accurate tool for the study of SIM properties of lanthanide complexes, especially considering that dynamical correlation plays a non-negligible role in the crystal-field splitting of high-lying Kramers doublets, as discussed recently by Haldar \textit{et al} \cite{Haldar2025}.

\begin{table}[H]
	\centering
	\caption{Excitation energies (in eV) of five 4f-5d excited states in \textbf{3Ce} obtained from all-electron (AE) and different DMET variants based CASSCF and NEVPT2 calculations.}
	\label{tab:3Ce}
	\resizebox{\textwidth}{!}{%
		\begin{tabular}{lccccccccc}
			\toprule
			Method
			& AO    & LO(IP) & LO(L\"owdin) & LO(IAO+PAO) & AO 	& LO(IP) & LO(L\"owdin) & LO(IAO+PAO) & AE \\
			\midrule
			$A$     & \multicolumn{4}{c}{Ce(104)} 	& \multicolumn{4}{c}{CeN$_6$(284)}	& \\
			\cmidrule(lr){2-5} \cmidrule(lr){6-9}
			$N_{\mathrm{EO}}$ 	& 273 & 209 & 209 & 143 & 453 & 452 & 453 & 351 & 662 \\
			\midrule
			\multicolumn{10}{l}{\textbf{CASSCF}} \\
			\midrule
			5d1 & 4.55 & 4.55 & 5.01 & 4.83 & 4.41 & 4.41 & 4.56 & 4.47 & 4.38 \\
			5d2 & 5.01 & 5.01 & 5.42 & 5.45 & 4.84 & 4.84 & 5.05 & 4.99 & 4.79 \\
			5d3 & 5.24 & 5.24 & 5.68 & 5.67 & 5.06 & 5.06 & 5.29 & 5.21 & 5.01 \\
			5d4 & 6.38 & 6.38 & 7.13 & 7.15 & 6.20 & 6.20 & 6.40 & 6.38 & 6.18 \\
			5d5 & 6.50 & 6.50 & 7.27 & 7.26 & 6.32 & 6.32 & 6.53 & 6.49 & 6.30 \\
			\hline
			MAE & 0.20 & 0.20 & 0.77 & 0.74 & 0.03 & 0.03 & 0.23 & 0.17 & \\
			MARE (\%) & 3.8 & 3.8 & 14.3 & 13.6 & 0.7 & 0.7 & 4.4 & 3.3 & \\
			\midrule
			\multicolumn{10}{l}{\textbf{NEVPT2}} \\
			\midrule
			5d1 & 4.26 & 4.26 & 4.56 & 4.79 & 3.80 & 3.80 & 4.16 & 3.94 & 3.59 \\
			5d2 & 4.75 & 4.76 & 4.97 & 5.41 & 4.24 & 4.24 & 4.67 & 4.46 & 3.99 \\
			5d3 & 4.99 & 5.00 & 5.23 & 5.63 & 4.47 & 4.47 & 4.91 & 4.68 & 4.21 \\
			5d4 & 6.14 & 6.14 & 6.73 & 7.13 & 5.59 & 5.59 & 5.99 & 5.83 & 5.39 \\
			5d5 & 6.26 & 6.27 & 6.87 & 7.24 & 5.71 & 5.71 & 6.12 & 5.94 & 5.51 \\
			\hline
			MAE & 0.74 & 0.75 & 1.13 & 1.50 & 0.22 & 0.22 & 0.63 & 0.43 & \\
			MARE (\%) & 16.8 & 16.9 & 25.0 & 33.3 & 5.1 & 5.1 & 14.3 & 9.7 & \\
			\bottomrule
	\end{tabular}}
\end{table}

\begin{table}[H]
	\centering
	\caption{Excitation energies (in eV) of five 4f-5d excited states in \textbf{4Ce} obtained from all-electron (AE) and different DMET variants based CASSCF and NEVPT2 calculations.}
	\label{tab:4Ce}
	\resizebox{\textwidth}{!}{%
		\begin{tabular}{lccccccccc}
			\toprule
			Method
			& AO & LO(IP) & LO(L\"owdin) & LO(IAO+PAO) 	& AO & LO(IP) & LO(L\"owdin) & LO(IAO+PAO)& AE \\
			\midrule
			$A$	& \multicolumn{4}{c}{Ce(104)} 	& \multicolumn{4}{c}{CeO$_6$(284)}	& \\
			\cmidrule(lr){2-5} \cmidrule(lr){6-9}
			$N_{\mathrm{EO}}$ 	& 305 & 209 & 209 & 142 & 591 & 569 & 569 & 352 & 878 \\
			\midrule
			\multicolumn{10}{l}{\textbf{CASSCF}} \\
			\midrule
			5d1 & 4.29 & 4.30 & 4.60 & 4.79 & 4.19 & 4.19 & 4.27 & 4.37 & 4.18 \\
			5d2 & 4.59 & 4.59 & 4.96 & 5.15 & 4.50 & 4.50 & 4.59 & 4.71 & 4.49 \\
			5d3 & 4.70 & 4.70 & 5.08 & 5.25 & 4.60 & 4.60 & 4.70 & 4.81 & 4.59 \\
			5d4 & 7.64 & 7.64 & 8.03 & 8.76 & 7.32 & 7.32 & 7.45 & 7.77 & 7.29 \\
			5d5 & 7.80 & 7.80 & 8.21 & 8.96 & 7.47 & 7.47 & 7.61 & 7.94 & 7.44 \\
			\hline
			MAE & 0.21 & 0.21 & 0.58 & 0.98 & 0.02 & 0.02 & 0.12 & 0.32 & \\
			MARE (\%) & 3.4 & 3.4 & 10.3 & 16.8 & 0.3 & 0.3 & 2.2 & 5.4 & \\
			\midrule
			\multicolumn{10}{l}{\textbf{NEVPT2}} \\
			\midrule
			5d1 & 3.88 & 3.88 & 4.18 & 4.78 & 3.72 & 3.72 & 3.84 & 4.03 & 3.69 \\
			5d2 & 4.20 & 4.20 & 4.58 & 5.16 & 4.04 & 4.04 & 4.17 & 4.39 & 4.01 \\
			5d3 & 4.30 & 4.31 & 4.71 & 5.26 & 4.14 & 4.14 & 4.27 & 4.48 & 4.11 \\
			5d4 & 7.30 & 7.31 & 7.66 & 8.71 & 6.75 & 6.75 & 6.93 & 7.30 & 6.64 \\
			5d5 & 7.46 & 7.46 & 7.83 & 8.91 & 6.89 & 6.89 & 7.08 & 7.45 & 6.77 \\
			\hline
			MAE & 0.39 & 0.39 & 0.75 & 1.52 & 0.07 & 0.07 & 0.21 & 0.49 & \\
			MARE (\%) & 7.0 & 7.1 & 14.6 & 29.8 & 1.2 & 1.2 & 4.2 & 9.6 & \\
			\bottomrule
		\end{tabular}%
	}
\end{table}

\subsection{4f-5d excitations in Ce(III) complexes}

We further consider excitation energies of \textbf{3Ce} and \textbf{4Ce}, which belong to the family of Ce(III) luminescent complexes that exhibit intriguing luminescence behavior arising from 4f--5d transitions \cite{Bunzli2013, WangL2022}. Because 5d orbitals in lanthanide elements are spatially more extended and interact more strongly with surrounding coordinating atoms than 4f orbitals, accurate description of 4f-5d excitations poses a more stringent test on the performance of embedding approaches like DMET.

In Table~\ref{tab:3Ce}, we report the excitation energies of 4f--5d transitions in \textbf{3Ce} obtained from all-electron and different DMET-based CASSCF and NEVPT2. Comparison of the CASSCF and NEVPT2 results shows that the inclusion of dynamical correlation significantly reduces the 4f--5d excitation energies by about 0.8 eV, consistent with the results reported by Sun \textit{et al}\cite{SunH2023}.

At both CASSCF and NEVPT2 levels, all results obtained from LO(IP)-DMET are essentially identical to those from AO-DMET even though the EO space of the former is considerably smaller than that of the latter when Ce is chosen as the impurity, and their agreement with all-electron results is significantly better than that from \LOlow~and LO(IAO+PAO)-DMET. To be more specific, at the CASSCF level, the MAEs of LO(IP)/AO-DMET results compared to all-electron ones are only 0.20 eV in the Ce-only case, and further decrease to 0.03 eV when treating \ce{CeN6} as the impurity, while the MAEs for the results from \LOlow~and LO(IAO+PAO)-DMET are 0.77 and 0.74 eV for the Ce-only impurity, and 0.23 and 0.17 eV for the \ce{CeN6} impurity, respectively. At the NEVPT2 level, MAEs from all DMET schemes compared to the all-electron treatment increase significantly due to the more delocalized nature of dynamical correlation that is more challenging to treat accurately by local correlation approaches like DMET, especially when Ce alone is chosen as the impurity. Nevertheless, LO(IP)/AO-DMET perform much better than \LOlow~and LO(IAO+PAO)-DMET. When treating \ce{CeN6} as the impurity, LO(IP)- and AO-DMET have nearly identical $N_{\rm EO}$ and lead to the MAE of only about 0.2 eV that is significantly smaller than the MAEs of \LOlow~and LO(IAO+PAO)-DMET (0.63 and 0.47 eV, respectively).

Table \ref{tab:4Ce} shows the results for \textbf{4Ce}, which exhibit trends similar to those discussed above for \textbf{3Ce}. LO(IP)- and AO-DMET again give nearly identical results at both the CASSCF and NEVPT2 levels although $N_{\rm EO}$ of the former is always smaller than that of the latter, and their MAEs are significantly smaller than those from \LOlow- and LO(IAO+PAO)-DMET. In particular, when treating Ce and its coordinate atoms (i.e. \ce{CeO6}) as the impurity, the MAE of LO(IP)/AO-DMET is only 0.02 eV at the CASSCF level, and slightly increases to 0.07 eV at the NEVPT2 level, which is remarkable considering the greatly reduced computational cost compared to the all-electron calculation.

The results for \textbf{3Ce} and \textbf{4Ce} shown in Tables \ref{tab:3Ce} and \ref{tab:4Ce} show some additional features that are noteworthy. Firstly, the results obtained from LO(IAO+PAO)-DMET become comparable to those from \LOlow-DMET even though \NEO~in the former is much smaller than that in the latter, in contrast to the trends observed for Dy-SIMs in the preceding section where the former performs much worse than the latter, indicating that the relative performances of the DMET schemes with different LOs are property dependent. Secondly, comparing the results for \textbf{3Ce} and \textbf{4Ce} with \ce{CeN6} or \ce{CeO6} as the impurity, one can see that although the number of LOs/AOs that define the impurity space is same in these two complexes, the number of EOs (equivalently, bath orbitals) in \textbf{4Ce} is significantly larger than that in \textbf{3Ce}. This phenomenon highlights an important general feature of the DMET approach. Although the upper bound for the number of bath orbitals in the LO-DMET formalism is equal to the number of LOs centered in the impurity plus the number of open-shell MOs, the actual number of bath orbitals generated by the mathematical procedure described in the preceding section depends on the interaction strength between the chosen impurity and its environment as embodied in the low-level reference wavefunction.

\section{CONCLUDING REMARKS}
 
To summarize, we have proposed an improved orthogonal partitioning scheme in the DMET framework that can achieve essentially the same accuracy of AO-DMET for strongly correlated systems while retaining the smaller EO space characteristics of the LO-DMET. We have presented a detailed mathematical proof showing that the EO space of the proposed LO(IP)-DMET forms a subspace of the EO space of AO-DMET with the same impurity choice, thereby providing insights into the relationship between these two methods. As indicated by extensive tests on four strongly correlated lanthanide complexes, at both CASSCF and NEVPT2 levels, LO(IP)- and AO-DMET yield results that are significantly closer to the AE references than those obtained with LO(L\"owdin) and LO(IAO+PAO)-DMET, suggesting their effectiveness in describing both static and dynamical correlations in local excitations. Remarkably, LO(IP)-DMET leads to essentially the same accuracy as AO-DMET but has a much smaller EO space when treating the lanthanide atom (Dy or Ce) as the impurity, thus offering a significant advantage in terms of computational efficiency. These results demonstrate the favorable properties and strong potential of the proposed IP partitioning scheme for DMET treatments of lanthanide complexes, particularly when an accurate description of excited state properties is required.

From a theoretical perspective, the present results, together with the explicitly proved mathematical relationship between LO(IP)- and AO-DMET, suggest that the accuracy improvement of AO-DMET for excited state properties observed in our previous work \cite{Ai2025b} is not merely a consequence of employing a non-orthogonal partition, but is more closely related to whether the partitioning scheme can provide an EO space that captures the local correlation involved in the excitation process. We speculate that the shared capability of LO(IP)- and AO-DMET in describing local correlation accurately arises from the fact that both methods retain the complete impurity space represented by AOs on the central lanthanide ion (and optionally its ligand atoms) in the embedded impurity space.

We conclude this article by discussing the limitations of our current approach and possible improvements in the future. In this work, LO(IP)- and AO-DMET yield nearly identical results, even though the EO space of LO(IP) is only a subspace of that of the latter. The essentially negligible contribution of the additional EOs generated from the non-orthogonal atomic orbitals based DMET to the description of both static and dynamical correlations associated with local electronic excitations remains to be understood from a more rigorous theoretical perspective, which will hopefully further elucidate the relationship between the two partitioning schemes and may provide a deeper understanding of the origin of favorable performance of LO(IP)-DMET.
In addition, although the IP scheme provides superior accuracy for local excitation energies compared with other commonly used partitioning schemes in DMET, whether an even better partitioning scheme exists remains an open question. We envision that a theoretically justified optimal partitioning scheme for local electronic excitations may be obtained by formulating this task as a constrained optimization problem. In this context, several previous studies have provided some conceptually relevant perspectives that may prove inspiring. Sun \textit{et al.} proposed an optimal treatment of quantum mechanics/molecular mechanics (QM/MM) boundaries \cite{Sun2014} by introducing exact link orbitals that fully capture QM effects across arbitrary QM/MM boundaries, in the sense that they ensure, in principle, zero truncation-induced error in the observables of QM region.
In the context of active-space-based quantum embedding, where a high-level solver is applied in a predefined active orbital space in the presence of a frozen-core potential, Lau \textit{et al.}\cite{Lau2024} compared the convergence behavior of different choices of active orbitals for CCSD excitation energies of defects in solids, including HF canonical molecular orbitals, local orbitals defined by regional embedding \cite{Lau2021}, and natural transition orbitals (NTOs) \cite{Martin2003}. The concept of quantum entanglement \cite{Peschel2012, Izsak2023, LiaoK2024} also provides a promising route for further developments of quantum embedding and has recently been employed in DMET bath-expansion strategies \cite{Nusspickel2022, Giordano2026}, which enables a systematically improvable description of structural and electronic properties of the total system. The integration of these methodological advances into the present framework may further improve the accuracy and robustness of DMET-based theoretical descriptions of local excitations in strongly correlated systems.

\section*{Data Availability}
The scripts and code used to generate and analyze the results, and the data underlying this study are available at [https://github.com/ccme-tmc/IPDMET-data].

\begin{acknowledgement}
This work is partly supported by Beijing Natural Science Foundation (Project Number 2252006) and National Natural Science Foundation of China (Project Number 12234001). We acknowledge the High-performance Computing Platform of Peking University for providing the computational facility.
\end{acknowledgement}

\begin{suppinfo}
Supporting Information available: (1) Mathematical proof that the EO space of LO(IP)-DMET is a subspace of that of AO-DMET; (2) Details of the construction of IAOs and PAOs.
\end{suppinfo}

\bibliography{Refs-IPDMET.bib}

\end{document}